\documentclass[a4paper,10pt]{article}
\usepackage[dvips]{graphicx}
\usepackage{amssymb,amsmath}
\numberwithin{equation}{section}

\begin{document}
\title{\bf{Integrable motion of two interacting curves,    spin systems  and the Manakov system}}
\author{ Akbota  Myrzakul\footnote{Email: akbota.myrzakul@gmail.com}   \, and    Ratbay Myrzakulov\footnote{Email: rmyrzakulov@gmail.com} \\ \textit{Eurasian International Center for Theoretical Physics and  Department of General } \\ \textit{ $\&$  Theoretical Physics, Eurasian National University, Astana 010008, Kazakhstan}}


\date{}
\maketitle
\begin{abstract}
Integrable spin systems   are an important subclass of integrable (soliton) nonlinear equations. They play important role in physics and mathematics. At present, many integrable spin systems were found and studied. They are related with the motion of 3-dimensional curves. In this paper, we  consider a  model of two  moving   interacting curves. Next, we find  its  integrable reduction related with some integrable coupled  spin system. Then we show that this integrable coupled spin system is equivalent to the famous Manakov system.
\end{abstract}
\vspace{2cm}
\section{Introduction}
 Among the  integrable systems, the integrable spin systems in 1+1 and 2+1 dimensions play an important role in physics and mathematics \cite{Gut}-\cite{myrzakulov-9535}. In physics, they  describe nonlinear dynamics of magnets. In differential geometry, they  can reproduce some integrable classes of curves and surfaces  \cite{myrzakulov-715}-\cite{myrzakulov-314}. The first and the most known representative of the  integrable spin  systems is the Heisenberg ferromagnetic equation (HFE)  which has a form
  \begin{eqnarray}
iA_{t}+\frac{1}{2}[A,A_{xx}]=0, \label{2.2} 
\end{eqnarray}
where ${\bf A}=(A_{1}, A_{2}, A_{3})$ is a unit spin vector, ${\bf A}^{2}=1$ and 
\begin{eqnarray}
A=\begin{pmatrix} A_{3}&A^{-}\\ 
A^{+}&-A_{3}\end{pmatrix}, \quad A^{2}=I=diag(1, 1), \quad A^{\pm}=A_{1}\pm i A_{2}. \label{2.2} 
\end{eqnarray}
The HFE is the Lakshmanan equivalent \cite{l77}  to the nonlinear Schr\"odinger equation 
 \begin{eqnarray}
iq_{t}+q_{xx}+2|q|^{2}q=0. \label{2.2} 
\end{eqnarray}
Also,  it is well-known that these equations are gauge equivalent to  each  other \cite{t77}.  At present,  many integrable and nonintegrable spin systems were identified (see e.g. Refs.\cite{Chen1}-\cite{akbota2} and references therein).  One of  such spin systems is the Myrzakulov-LIII  equation or shortly,  the M-LIII equation\footnote{Here LIII $\equiv$ 53 so that  M-LIII $\equiv$ M-53 and the M-LIII equation $\equiv$ the M-53 equation.}. The M-LIII equation reads as  \cite{myrzakulov-391}-\cite{M35}  
  \begin{eqnarray}
iA_{t}+\frac{1}{2}[A,A_{xx}]+iuA_{x}=0, \label{2.2} 
\end{eqnarray}
where $u=u(t,x)$ is some real  function (potential). The modified (inhomogeneous)  M-LIII equation looks like
  \begin{eqnarray}
iA_{t}+\frac{1}{2}[A,A_{xx}]+iuA_{x}+F=0, \label{2.2} 
\end{eqnarray}
where $F$ is a matrix function. In this paper, we study the two-layer ("two-component") generalization of the modified M-LIII equation (1.5) or in short, the coupled M-LIII equation. 

The paper is organized as follows.  In Sec. 2, the coupled M-LIII  equation is  introduced.   In Sec. 3, we derived the Lakshmanan equivalent counterpart  of the M-LIII equation, namely, the Manakov system.   In Sec. 4, we present two types of Lax representations of the coupled M-LIII equation. In Sec.  5, the relation between the solutions of the Manakov system and the M-LIII equation is  established. The gauge equivalent counterpart of the Manakov system is presented in Sec.  6. The relation between solutions of the coupled M-LIII equation and the $\Gamma$-spin system is considered in Sec. 7. At last, Sec. 8 is devoted to Conclusions.


\section{The coupled  M-LIII equation}

 Consider two spin vectors ${\bf A}=(A_{1}, A_{2}, A_{3})$ and  ${\bf B}=(B_{1}, B_{2}, B_{3})$, where ${\bf A}^{2}={\bf B}^{2}=1$. Let these spin vectors satisfy the  coupled  Myrzakulov-LIII equation or the 2-layer Myrzakulov-LIII equation  of the form \cite{akbota1}-\cite{akbota8}
\begin{eqnarray}
iA_{t}+\frac{1}{2}[A,A_{xx}]+iu_{1}A_{x}+F&=&0,\\
iB_{t}+\frac{1}{2}[B,B_{xx}]+iu_{2}B_{x}+E&=&0. \label{2.2} 
\end{eqnarray}
Here $u_{k}$
are real functions, $F$ and $E$  are matrix functions, $B$ is the matrix form of the ${\bf B}$  (second  spin vector)
\begin{eqnarray}
B=\begin{pmatrix} B_{3}&B^{-}\\ 
B^{+}&-B_{3}\end{pmatrix}, \quad F=\begin{pmatrix} F_{3}&F^{-}\\ 
F^{+}&-F_{3}\end{pmatrix}, \quad
E=\begin{pmatrix} E_{3}&E^{-}\\ 
E^{+}&-E_{3}\end{pmatrix},  \label{2.2} 
\end{eqnarray}
where $B^{\pm}=B_{1}\pm iB_{2}, \quad B^{2}=I, \quad F^{\pm}=F_{1}\pm iF_{2},\quad E^{\pm}=E_{1}\pm iE_{2}.$ 
We now introduce two complex functions $u$ and $v$ as
\begin{eqnarray}
u=\frac{A^{+}}{1+A_{3}},\quad v=\frac{B^{+}}{1+B_{3}}.\label{2.16}
\end{eqnarray}
Then these functions satisfy the  following set of equations
 \begin{eqnarray}
iu_{t}-u_{xx}+\frac{2u^{*}u_{x}^{2}}{1+|u|^{2}}&=&F^{\prime},\label{3.666}\\
iv_{t}-v_{xx}+\frac{2v^{*}v_{x}^{2}}{1+|v|^{2}}&=&E^{\prime},\label{2.16}
\end{eqnarray}
where $F^{\prime}$ and $E^{\prime}$ are some complex functions
\begin{eqnarray}
F^{\prime}&=&F^{\prime}(u,v,u_{x}, v_{x}, ...),\label{3.666}\\
E^{\prime}&=&E^{\prime}(u,v,u_{x}, v_{x}, ...).\label{2.16}
\end{eqnarray}
In this paper, we assume that $F$ and $E$ have the form
\begin{eqnarray}
F=v_{1}[\sigma_{3},A],\quad E=v_{2}[\sigma_{3},B], \label{2.2} 
\end{eqnarray}
where  $v_{j}$ are some real functions (potentials). 
Then the  coupled M-LIII equation (2.1)-(2.2) takes the form
\begin{eqnarray}
iA_{t}+\frac{1}{2}[A,A_{xx}]+iu_{1}A_{x}+v_{1}[\sigma_{3},A]&=&0,\\
iB_{t}+\frac{1}{2}[B,B_{xx}]+iu_{2}B_{x}+v_{2}[\sigma_{3},B]&=&0. \label{2.2} 
\end{eqnarray}
Here $u_{j}$ and $v_{j}$ are coupling potentials and have the following forms
\begin{eqnarray}
u_{1}&=&\frac{i({\bar Z}B^{-}-ZB^{+})}{W(1+B_{3})},\label{3.666}\\
v_{1}&=&-\frac{|Z|^{2}}{2W(1+A_{3})^{2}},\label{2.16}\\
u_{2}&=&\frac{i({\bar R}A^{-}-RA^{+})(1+B_{3})}{W(1+A_{3})^{2}},\label{3.666}\\
v_{2}&=&-\frac{|R|^{2}}{2W(1+A_{3})^{3}},\label{2.16}
\end{eqnarray} 
where\begin{eqnarray}
W&=&2+\frac{(1+A_{3})(1-B_{3})}{1+B_{3}}, \label{2.2} \\
R&=&WA^{-}_{x}-MA^{-}, \label{2.2} \\
Z&=&W[(1+A_{3})(1+B_{3})^{-1}B^{-} ]_{x}-M[(1+A_{3})(1+B_{3})^{-1} B^{-}],\\
M&=&A_{3x}+\frac{A^{+}A^{-}_{x}}{1+A_{3}}+\frac{A_{3x}(1-B_{3})}{1+B_{3}}+\frac{(1+A_{3})B^{+}B^{-}_{x}}{(1+B_{3})^{2}}-\frac{(1+A_{3})(1-B_{3})B_{3x}}{(1+B_{3})^{2}}.
\end{eqnarray}
In components,  the 2-layer  M-LIII equation (2.10)-(2.11) reads as
\begin{eqnarray}
iA_{t}^{+}+(A^{+}A_{3xx}-A^{+}_{xx}A_{3})+iu_{1}A^{+}_{x}-2v_{1}A^{+}&=&0,\\
iA_{t}^{-}-(A^{-}A_{3xx}-A^{-}_{xx}A_{3})+iu_{1}A_{x}^{-}+2v_{1}A^{-}&=&0,\\
iA_{3t}+\frac{1}{2}(A^{-}A^{+}_{xx}-A^{-}_{xx}A^{+})+iu_{1}A_{3x}&=&0,\\
iB_{t}^{+}+(B^{+}B_{3xx}-B^{+}_{xx}B_{3})+iu_{2}B_{x}^{+}-2v_{2}B^{+}&=&0, \label{2.2} \\
iB_{t}^{-}-(B^{-}B_{3xx}-B^{-}_{xx}B_{3})+iu_{2}B_{x}^{-}+2v_{2}B^{-}&=&0, \label{2.2} \\
iB_{3t}+\frac{1}{2}(B^{-}B^{+}_{xx}-B^{-}_{xx}B^{+})+iu_{2}B_{3x}&=&0 \label{2.2} 
\end{eqnarray}
or
\begin{eqnarray}
A_{1t}+A_{2}A_{3xx}-A_{2xx}A_{3}+u_{1}A_{1x}-2v_{1}A_{2}&=&0,\\
A_{2t}+A_{3}A_{1xx}-A_{3xx}A_{1}+u_{1}A_{2x}-2v_{1}A_{1}&=&0,\\
A_{3t}+A_{1}A_{2xx}-A_{1xx}A_{2}+u_{1}A_{3x}&=&0,\\
B_{1t}+B_{2}B_{3xx}-B_{2xx}B_{3}+u_{2}B_{1x}-2v_{2}B_{2}&=&0,\\
B_{2t}+B_{3}B_{1xx}-B_{3xx}B_{1}+u_{2}B_{2x}-2v_{2}B_{1}&=&0,\\
B_{3t}+B_{1}B_{2xx}-B_{1xx}B_{2}+u_{2}B_{3x}&=&0. \label{2.2} 
\end{eqnarray}
\section{Lakshmanan equivalent counterpart of the coupled M-LIII equation}
In this section, we present the Lakshmanan equivalent counterpart of the coupled  M-LIII equation (2.10)-(2.11). To do that, let us rewrite the 2-layer   M-LIII equation (2.10)-(2.11) in the vector form as \cite{akbota1}-\cite{akbota8}
\begin{eqnarray}
{\bf A}_{t}+{\bf A}\wedge {\bf A}_{xx}+u_{1}{\bf A}_{x}+2v_{1}{\bf H}\wedge {\bf A}&=&0,\\
{\bf B}_{t}+{\bf B}\wedge {\bf B}_{xx}+u_{2}{\bf B}_{x}+2v_{2}{\bf H}\wedge {\bf B}&=&0, \label{2.2} 
\end{eqnarray} 
where ${\bf H}=(0,0,1)$ is the constant magnetic field. 
Now we consider  two interacting 3-dimensional curves in  $R^{n}$. These curves are given by the  following two basic vectors 
${\bf e}_{k}$ and ${\bf l}_{k}$. The  motion of these   curves is defined by the following  
equations 
\begin{eqnarray}
\left ( \begin{array}{ccc}
{\bf  e}_{1} \\
{\bf  e}_{2} \\
{\bf  e}_{3}
\end{array} \right)_{x} = C
\left ( \begin{array}{ccc}
{\bf  e}_{1} \\
{\bf  e}_{2} \\
{\bf  e}_{3}
\end{array} \right),\quad
\left ( \begin{array}{ccc}
{\bf  e}_{1} \\
{\bf  e}_{2} \\
{\bf  e}_{3}
\end{array} \right)_{t} = D
\left ( \begin{array}{ccc}
{\bf  e}_{1} \\
{\bf  e}_{2} \\
{\bf  e}_{3}
\end{array} \right) \label{2.1} 
\end{eqnarray}
and
\begin{eqnarray}
\left ( \begin{array}{ccc}
{\bf  l}_{1} \\
{\bf  l}_{2} \\
{\bf  l}_{3}
\end{array} \right)_{x} = L
\left ( \begin{array}{ccc}
{\bf  l}_{1} \\
{\bf  l}_{2} \\
{\bf  l}_{3}
\end{array} \right),\quad
\left ( \begin{array}{ccc}
{\bf  l}_{1} \\
{\bf  l}_{2} \\
{\bf  l}_{3}
\end{array} \right)_{t} = N
\left ( \begin{array}{ccc}
{\bf  l}_{1} \\
{\bf  l}_{2} \\
{\bf  l}_{3}
\end{array} \right). \label{2.1} 
\end{eqnarray}
Here ${\bf e}_{1}, {\bf e}_{2}$ and ${\bf e}_{3}$ are the unit tangent, normal 
and binormal vectors respectively to the first curve,  ${\bf l}_{1}, {\bf l}_{2}$ and ${\bf l}_{3}$ are the unit tangent, normal 
and binormal vectors respectively to the second curve,  $x$ is the arclength 
parametrising these both  curves. The matrices $C, D, L, N$ are given by
\begin{eqnarray}
C =
\left ( \begin{array}{ccc}
0   & k_{1}     & 0 \\
-k_{1} & 0     & \tau_{1}  \\
0   & -\tau_{1} & 0
\end{array} \right) ,\quad
G =
\left ( \begin{array}{ccc}
0       & \omega_{3}  & -\omega_{2} \\
-\omega_{3} & 0      & \omega_{1} \\
\omega_{2}  & -\omega_{1} & 0
\end{array} \right),\label{2.2} 
\end{eqnarray}
\begin{eqnarray}
L =
\left ( \begin{array}{ccc}
0   & k_{2}     & 0 \\
-k_{2} & 0     & \tau_{2}  \\
0   & -\tau_{2} & 0
\end{array} \right) ,\quad
N =
\left ( \begin{array}{ccc}
0       & \theta_{3}  & -\theta_{2} \\
-\theta_{3} & 0      & \theta_{1} \\
\theta_{2}  & -\theta_{1} & 0
\end{array} \right).\label{2.2} 
\end{eqnarray}
For the   curvatures and torsions of  curves we obtain
\begin{eqnarray}
k_{1} & = & \sqrt{{\bf e}_{1x}^{2}},\quad \tau_{1}= \frac{{\bf e}_{1}\cdot ({\bf e}_{1x} \wedge {\bf e}_{1xx})}{{\bf e}_{1x}^{2}}, \\
k_{2} & = &\sqrt{ {\bf l}_{1x}^{2}},\quad \tau_{2}= \frac{{\bf l}_{1}\cdot ({\bf l}_{1x} \wedge {\bf l}_{1xx})}{{\bf l}_{1x}^{2}}.        \label{2.3}
\end{eqnarray}
The  equations (3.3) and (3.4) are compatible if 
\begin{eqnarray}
C_t - G_x + [C, G] &=& 0,\label{2.4} \\
L_t - N_x + [L, N] &=& 0,\label{2.4} 
\end{eqnarray}
respectively. In elements these equations take the form
 \begin{eqnarray}
k_{1t}    & = & \omega_{3x} + \tau_{1} \omega_2, \label{2.5} \\ 
\tau_{1t}      & = & \omega_{1x} - k_{1}\omega_2, \\ \label{2.6} 
\omega_{2x} & = & \tau_{1} \omega_3-k_{1} \omega_1 \label{2.7} 
\end{eqnarray}
and
 \begin{eqnarray}
k_{2t}    & = & \theta_{3x} + \tau_{2} \theta_2, \label{2.5} \\ 
\tau_{2t}      & = & \theta_{1x} - k_{2}\theta_2, \\ \label{2.6} 
\theta_{2x} & = & \tau_{2} \theta_3-k_{2} \theta_1, \label{2.7} 
\end{eqnarray}
respectively. 
Our next step is  the following identifications:
 \begin{eqnarray}
{\bf A}\equiv {\bf e}_{1}, \quad {\bf B}\equiv {\bf l}_{1}. \label{2.7} 
\end{eqnarray}
We also assume that
\begin{eqnarray}
{\bf F}=F_{1}{\bf e}_{1}+F_{2}{\bf e}_{2}+F_{3}{\bf e}_{3}, \quad {\bf E}=E_{1}{\bf l}_{1}+E_{2}{\bf l}_{2}+E_{3}{\bf l}_{3}, \label{2.7} 
\end{eqnarray}
where
\begin{eqnarray}
{\bf F}=2v_{1}{\bf H}\wedge {\bf A}, \quad {\bf E}=2v_{2}{\bf H}\wedge {\bf B}. \label{2.7} 
\end{eqnarray}
Then we obtain
\begin{eqnarray}
k_{1} & = & \sqrt{{\bf A}_{x}^{2}},\\ 
\tau_{1}&=&  \frac{{\bf A}\cdot ({\bf A}_{x} \wedge {\bf A}_{xx})}{{\bf A}_{x}^{2}}, \\
k_{2} & = & \sqrt{{\bf B}_{x}^{2}},\\
\tau_{2}&=&  \frac{{\bf B}\cdot ({\bf B}_{x} \wedge {\bf B}_{xx})}{ {\bf B}_{x}^{2}}        \label{2.3}
\end{eqnarray}
and
\begin{eqnarray}
\omega_{1} & = & -\frac{k_{1xx}+F_{2}\tau_{1}+F_{3x}}{k_{1}}+(\tau_{1}-u_{1})\tau_{1},\\ 
\omega_{2}&=& k_{1x}+F_{3}, \\
\omega_{3} & = &k_{1}(\tau_{1}-u_{1})-F_{2},\\
\theta_{1} & = & -\frac{k_{2xx}+E_{2}\tau_{2}+E_{3x}}{k_{2}}+(\tau_{2}-u_{2})\tau_{2},\\ 
\theta_{2}&=& k_{2x}+E_{3}, \\
\theta_{3} & = &k_{2}(\tau_{2}-u_{2})-E_{2}        \label{2.3}
\end{eqnarray}
with
\begin{eqnarray}
F_{1}=E_{1}=0.   \label{2.3}
\end{eqnarray}
We now can  write the equations for $k_{j}$ and $\tau_{j}$. They look like
 \begin{eqnarray}
k_{1t}&=&2k_{1x}\tau_{1}+k_{1}\tau_{1x}-(u_{1}k_{1})_{x}-F_{2x}+F_{3}\tau_{1}, \label{2.5} \\ 
\tau_{1t}&=&\left[-\frac{k_{1xx}+F_{2}\tau_{1}+F_{3x}}{k_{1}}+(\tau_{1}-u_{1})\tau_{1}-\frac{1}{2}k_{1}^{2}\right]_{x}-F_{3}k_{1},  \label{2.7} \\
k_{2t}&=&2k_{2x}\tau_{2}+k_{2}\tau_{2x}-(u_{2}k_{2})_{x}-E_{2x}+E_{3}\tau_{2}, \label{2.5} \\ 
\tau_{2t}&=&\left[-\frac{k_{2xx}+E_{2}\tau_{2}+E_{3x}}{k_{2}}+(\tau_{2}-u_{2})\tau_{2}-\frac{1}{2}k_{2}^{2}\right]_{x}-E_{3}k_{2}. \label{2.7} 
\end{eqnarray}
Let us now  introduce new four  real functions $\alpha_{j}$ and $\beta_{j}$ as
\begin{eqnarray}
\alpha_{1}&=&0.5k_{1}\sqrt{1+ \zeta_{1}},\\ 
\beta_{1}&=&\tau_{1}(1+ \xi_{1}),\\ 
\alpha_{2}&=&0.5k_{2}\sqrt{1+ \zeta_{2}},\\
\beta_{2}&=&\tau_{2}(1+ \xi_{2}), \label{2.2} 
\end{eqnarray}
where
\begin{eqnarray}
\zeta_{1}&=&\frac{2|WA^{-}_{x}-MA^{-}|^{2}}{W^{2}(1+A_{3})^{2}{\bf A}^{2}_{x}}-1,\\ 
\zeta_{2}&=&\frac{2|W[(1+A_{3})(1+B_{3})^{-1}B^{-} ]_{x}-M[(1+A_{3})(1+B_{3})^{-1} B^{-}]|^{2}}{W^{2}(1+A_{3})^{2}{\bf B}^{2}_{x}}-1,\label{2.2} \\
\xi_{1}&=&\frac{\bar{R}_{x}R-\bar{R}R_{x}-4i|R|^{2}\nu_{x}}{2i\alpha_{1}^{2}W^{2}(1+A_{3})^{2}\tau_{1}}-1,\\ 
\xi_{2}&=&\frac{\bar{Z}_{x}Z-\bar{Z}Z_{x}-4i|Z|^{2}\nu_{x}}{2i\alpha_{2}^{2}W^{2}(1+A_{3})^{2}\tau_{2}}-1. \label{2.2} 
\end{eqnarray}
Here 
\begin{eqnarray}
\nu&=& \partial^{-1}_{x}\left[\frac{A_{1}A_{2x}-A_{1x}A_{2}}{(1+A_{3})W}-\frac{(1+A_{3})(B_{1x}B_{2}-B_{1}B_{2x})}{(1+B_{3})^{2}W}\right].\label{2.2} 
\end{eqnarray}

 We now ready to  write the equations for the functions $\alpha_{i}$ and $\beta_{j}$. They   satisfy the following four equations
\begin{eqnarray}
\alpha_{1t}-2\alpha_{1x}\beta_{1}-\alpha_{1}\beta_{1x}&=&0,\\ 
\beta_{1t}+\left[\frac{\alpha_{1xx}}{\alpha_{1}}-\beta_{1}^{2}+2(\alpha_{1}^{2}+\alpha_{2}^{2})\right]_{x}&=&0, \\ 
\alpha_{2t}-2\alpha_{2x}\beta_{2}-\alpha_{2}\beta_{2x}&=&0,\\ 
\beta_{2t}+\left[\frac{\alpha_{2xx}}{\alpha_{2}}-\beta_{2}^{2}+2(\alpha_{1}^{2}+\alpha_{2}^{2})\right]_{x}&=&0. \label{2.2} 
\end{eqnarray}
Let us now  we introduce new two complex functions using by the A-transformation. The A-transformation reads  as (see e.g. \cite{akbota1}-\cite{akbota8}) 
\begin{eqnarray}
q_{1}&=&\alpha_{1}e^{-i\partial^{-1}_{x}\beta_{1}},\\ q_{2}&=&\alpha_{2}e^{-i\partial^{-1}_{x}\beta_{2}}. \label{2.2} 
\end{eqnarray}
Sometime we use the following explicit form of the A-transformation
\begin{eqnarray}
q_{1}&=&0.5k_{1}\sqrt{1+ \zeta_{1}}e^{-i\partial^{-1}_{x}[\tau_{1}(1+ \xi_{1})]},\\ q_{2}&=&0.5k_{2}\sqrt{1+ \zeta_{2}}e^{-i\partial^{-1}_{x}[\tau_{2}(1+ \xi_{2})]}. \label{2.2} 
\end{eqnarray}
 It is not difficult to verify that these new complex functions $q_{j}$ satisfy the following  system of equations (see e.g. \cite{Kostov})
\begin{eqnarray}
iq_{1t}+q_{1xx}+2(|q_{1}|^{2}+|q_{2}|^{2})q_{1}&=&0,\label{3.21}\\
 iq_{2t}+q_{2xx}+2(|q_{1}|^{2}+|q_{2}|^{2})q_{2}&=&0.\label{3.23}
\end{eqnarray}
It is nothing but  the Manakov system. So we proved that the Manakov system  (3.52)-(3.53) is the Lakshmanan equivalent counterpart of the 2-layer M-LIII equation  (2.10)-(2.11) or in the vector form (3.1)-(3.2). Finally we note that if $\zeta_{j}=\xi_{j}=0$ then  the A-transformation (3.48)-(3.49) or (3.50)-(3.51) reduces to  the Hasimoto transformation
 \begin{eqnarray}
q_{1}&=&0.5\kappa_{1}e^{-i\partial^{-1}_{x}\tau_{1}},\\ q_{2}&=&0.5\kappa_{2}e^{-i\partial^{-1}_{x}\tau_{2}}. \label{2.2} 
\end{eqnarray}

\section{Lax representation of the coupled M-LIII equation hierarchy}
In the previous section we have shown that the coupled M-LIII equation is the Lakshmanan equivalent to the Manakov system. This means that the coupled M-LIII equation is integrable by the IST method since its equivalent counterpart - the Manakov system is integrable. In turn, it means that  the M-LIII equation admits all ingredients of integrable systems like Lax representation (LR), infinite number of commuting integrals of motion, n-soliton solutions etc. Below  we present two possible versions of the LR for the M-LIII equation hierarchy.  
\subsection{LR type - I}
The first type of LR  for the coupled M-LIII equation hierarchy reads as
\begin{eqnarray}
Y_{x}&=&-i\lambda PY,\label{2.1}\\
Y_{t}&=&\sum_{j=1}^{N}\lambda^{j}V_{j} Y, \label{2.2} 
\end{eqnarray}
where $\lambda$ is a spectral parameter and 
\begin{eqnarray}
P =\frac{1}{2+K}
\left ( \begin{array}{ccc}
2A_{3}-K   & 2A^{-}     & \frac{2(1+A_{3})B^{-}}{1+B_{3}} \\
2A^{+} &   -(2A_{3}+K)&\frac{2A^{+}B^{-}}{1+B_{3}} \\
\frac{2(1+A_{3})B^{+}}{1+B_{3}}   & \frac{2A^{-}B^{+}}{1+B_{3}} & K-2
\end{array} \right),\label{2.2} 
\end{eqnarray}
with $K= (1+A_{3})(1-B_{3})(1+B_{3})^{-1}$.
The compatibility condition of this system gives the coupled M-LIII equation hierarchy. As the particular example, let us consider the case when $N=2$. Then the set of equations (4.1)-(4.2) takes the form
\begin{eqnarray}
Y_{x}&=&-i\lambda PY,\label{2.1}\\
Y_{t}&=&\left(\lambda^{2}V_{2}+\lambda V_{1}\right)Y, \label{2.2} 
\end{eqnarray}
where  
\begin{eqnarray}
V_{2}=-2iP, \quad V_{1}=PP_{x}.\label{2.2} 
\end{eqnarray}

The compatibility condition of the equations (4.4)-(4.5) gives  the 2-layer M-LIII equation (2.10)-(2.11) or (3.1)-(3.2) that is same.
\subsection{LR type - II}
The second  type of LR  for the coupled M-LIII equation hierarchy can be written in the following form
\begin{eqnarray}
Y_{x}&=&-i\lambda QY,\label{2.1}\\
Y_{t}&=&\sum_{j=1}^{N}\lambda^{j}W_{j} Y. \label{2.2} 
\end{eqnarray}
Here  
\begin{eqnarray}
Q =Q_{1}+Q_{2},\label{2.2} 
\end{eqnarray}
where
\begin{eqnarray}
Q_{1}=\left ( \begin{array}{cccc}
0   & A_{1}     & A_{2}& A_{3} \\
-A_{1} &  0&A_{3}&-A_{2} \\
-A_{2} &   -A_{3}& 0&A_{1} \\
-A_{3}   & A_{2}&-A_{1} &0
\end{array} \right),\quad Q_{2}=\left ( \begin{array}{cccc}
0   & B_{1}     & B_{2}& -B_{3} \\
-B_{1} &  0&B_{3}&B_{2} \\
-B_{2} &   -B_{3}& 0&-B_{1} \\
B_{3}   & -B_{2}&B_{1} &0
\end{array} \right).\label{2.2} 
\end{eqnarray} 
From the compatibility condition of the set of equations (4.7)-(4.8) $Y_{xt}=Y_{tx}$ we obtain  the coupled M-LIII equation hierarchy.  

\section{Relation between solutions of the coupled M-LIII equation and the Manakov system}

Let $A_{j}$ and $B_{j}$ be  the solution of the coupled M-LIII equation (2.10)-(2.11). Then the solution of the Manakov system  (3.52)-(3.53) is  given by 
\begin{eqnarray}
q_{1}&=&\frac{Re^{2i\nu}}{W(1+A_{3})},\\ 
q_{2}&=&\frac{Ze^{2i\nu}}{W(1+A_{3})}.\label{2.2} 
\end{eqnarray}
\section{Gauge equivalence between the $\Gamma$-spin system and  the Manakov system}
Above, we have proved that the coupled M-LIII equation (2.10)-(2.11) and the Manakov system  (3.52)-(3.53) is the Lakshmanan equivalent to each other. 
In this section, we want to present the another (gauge) equivalent counterpart of the Manakov system. 
It is well-known that the Lax representation of the Manakov  equation  (3.52)-(3.53) has the form (see e.g. \cite{Kostov})
\begin{eqnarray}
\Phi_{x}&=&U\Phi,\label{2.1}\\
\Phi_{t}&=&V\Phi. \label{2.2} 
\end{eqnarray}
Here
\begin{eqnarray}
U=-i\lambda \Sigma+U_{0}, \quad V=-2i\lambda^{2}\Sigma+2\lambda U_{0}+V_{0} \label{2.2} 
\end{eqnarray}
with
\begin{eqnarray}
\Sigma =
\left(\begin{array}{ccc}
1   & 0     & 0 \\
0 & -1    & 0  \\
0   & 0 & -1
\end{array}\right),\, U_{0} =
\left ( \begin{array}{ccc}
0       & q_{1}  & q_{2} \\
-\bar{q}_{1} & 0      & 0 \\
-\bar{q}_{2}  & 0 & 0
\end{array} \right), \,   V_{0} =
i\left ( \begin{array}{ccc}
|q_{1}|^{2}+|q_{2}|^{2}      & q_{1x}  & q_{2x} \\
\bar{q}_{1x} & - |q_{1}|^{2}     & -\bar{q}_{1}q_{2} \\
\bar{q}_{2x}  & -\bar{q}_{2}q_{1} & -|q_{2}|^{2}
\end{array}\right).\label{2.2} 
\end{eqnarray}
Let us now  consider the gauge transformation
\begin{eqnarray}
\Psi=g^{-1}\Phi, \quad g=\Phi_{\lambda=0}. \label{2.2} 
\end{eqnarray}
Then $\Psi$ obeys the equations
\begin{eqnarray}
\Psi_{x}&=&U^{\prime}\Psi,\label{2.1}\\
\Psi_{t}&=&V^{\prime}\Psi, \label{2.2} 
\end{eqnarray}
where
\begin{eqnarray}
U^{\prime}=-i\lambda\Gamma, \quad V^{\prime}=-2i\lambda^{2}\Gamma+\frac{1}{2}\lambda[\Gamma, \Gamma_{x}]. \label{2.2} 
\end{eqnarray}
Here
\begin{eqnarray}
\Gamma=g^{-1}\Sigma g, \quad \Gamma^{2}=I \label{2.2} 
\end{eqnarray}
and
\begin{eqnarray}
\Gamma =
\left ( \begin{array}{ccc}
\Gamma_{11}  & \Gamma_{12}    &\Gamma_{13} \\
\Gamma_{21} &   \Gamma_{22} &\Gamma_{23} \\
\Gamma_{31}   & \Gamma_{32} & \Gamma_{33}
\end{array} \right).\label{2.2} 
\end{eqnarray}
Elements of the $\Gamma$ matrix satisfy some restrictions 
\begin{eqnarray}
\Gamma_{33}=-(1
+\Gamma_{11}+\Gamma_{22}), \quad \Gamma_{ij}=\bar{\Gamma}_{ji},\label{2.2} 
\end{eqnarray}
and
\begin{eqnarray}
\Gamma_{ik}\Gamma_{kj}+\Gamma_{i(k+1)}\Gamma_{(k+1)i}+\Gamma_{i(k+2)}\Gamma_{(k+2)i}&=&0, (i\neq k\neq j),\\ \label{2.2} 
\Gamma_{ik}\Gamma_{ki}+\Gamma_{i(k+1)}\Gamma_{(k+1)i}+\Gamma_{i(k+2)}\Gamma_{(k+2)i}&=&1.
\end{eqnarray}
The compatibility condition of the equations (5.6)-(5.7) gives
\begin{eqnarray}
i\Gamma_{t}+\frac{1}{2}[\Gamma, \Gamma_{xx}]=0. \label{2.2} 
\end{eqnarray}
We call this equation  -  the $\Gamma$-spin system. Thus the $\Gamma$-spin system (6.14) is the gauge equivalent counterpart of the Manakov system  (3.52)-(3.53). It is the  well-known result (see e.g. \cite{Kostov}).  In terms of elements,  the $\Gamma$-spin system (6.14)  reads as 
\begin{eqnarray}
i\Gamma_{11t}+\frac{1}{2}(\Gamma_{12}\Gamma_{21xx}+\Gamma_{13}\Gamma_{31xx}-\Gamma_{12xx}\Gamma_{21}-\Gamma_{13xx}\Gamma_{31})&=&0,\label{2.1}\\
i\Gamma_{12t}+\frac{1}{2}(\Gamma_{11}\Gamma_{12xx}+\Gamma_{12}\Gamma_{22xx}+\Gamma_{13}\Gamma_{32xx}-\Gamma_{11xx}\Gamma_{12}-\Gamma_{12xx}\Gamma_{22}-\Gamma_{13xx}\Gamma_{32})&=&0,\label{2.1}\\
i\Gamma_{13t}+\frac{1}{2}(\Gamma_{11}\Gamma_{13xx}+\Gamma_{12}\Gamma_{23xx}+\Gamma_{13}\Gamma_{33xx}-\Gamma_{11xx}\Gamma_{13}-\Gamma_{12xx}\Gamma_{23}-\Gamma_{11xx}\Gamma_{13})&=&0,\label{2.1}\\
i\Gamma_{21t}+\frac{1}{2}(\Gamma_{21}\Gamma_{11xx}+\Gamma_{22}\Gamma_{21xx}+\Gamma_{23}\Gamma_{31xx}-\Gamma_{21xx}\Gamma_{11}-\Gamma_{22xx}\Gamma_{21}-\Gamma_{23xx}\Gamma_{31})&=&0,\label{2.1}\\
i\Gamma_{22t}+\frac{1}{2}(\Gamma_{21}\Gamma_{12xx}+\Gamma_{23}\Gamma_{32xx}-\Gamma_{21xx}\Gamma_{12}-\Gamma_{23xx}\Gamma_{32})&=&0,\label{2.1}\\
i\Gamma_{23t}+\frac{1}{2}(\Gamma_{21}\Gamma_{13xx}+\Gamma_{22}\Gamma_{23xx}+\Gamma_{23}\Gamma_{33xx}-\Gamma_{21xx}\Gamma_{13}-\Gamma_{22xx}\Gamma_{23}-\Gamma_{23xx}\Gamma_{33})&=&0,\label{2.1}\\
i\Gamma_{31t}+\frac{1}{2}(\Gamma_{31}\Gamma_{11xx}+\Gamma_{32}\Gamma_{21xx}+\Gamma_{33}\Gamma_{31xx}-\Gamma_{31xx}\Gamma_{11}-\Gamma_{32xx}\Gamma_{21}-\Gamma_{33xx}\Gamma_{31})&=&0,\label{2.1}\\
i\Gamma_{32t}+\frac{1}{2}(\Gamma_{31}\Gamma_{12xx}+\Gamma_{32}\Gamma_{22xx}+\Gamma_{33}\Gamma_{32xx}-\Gamma_{31xx}\Gamma_{12}-\Gamma_{32xx}\Gamma_{22}-\Gamma_{33xx}\Gamma_{32})&=&0,\label{2.1}\\
i\Gamma_{33t}+\frac{1}{2}(\Gamma_{31}\Gamma_{13xx}+\Gamma_{32}\Gamma_{23xx}-\Gamma_{31xx}\Gamma_{13}-\Gamma_{32xx}\Gamma_{23})&=&0. \label{2.2} 
\end{eqnarray}

\section{Relation between solutions of the coupled M-LIII equation and the $\Gamma$-spin system}
In the previous sections we have shown that to  the one and  same set of equations - the  Manakov system  (3.52)-(3.53),  correspond two spin systems: the coupled M-LIII equation (3.1)-(3.2) and the $\Gamma$-spin system (6.14). It tells us that between these two spin systems there must be some exact relation/correspondence. In other words, the 2-layer  M-LIII equation (2.10)-(2.11) and the $\Gamma$-spin system  (6.14) are equivalent  to each  other by some exact transformations. Below we will present these  transformations.
\subsection{Direct  M-transformation}

According to the M-transformation,  in terms of the spin vectors ${\bf A}$ and ${\bf B}$, the elements of the $\Gamma$-spin system are expressed as
\begin{eqnarray}
\Gamma =\frac{1}{2+K}
\left ( \begin{array}{ccc}
2A_{3}-K   & 2A^{-}     & \frac{2(1+A_{3})B^{-}}{1+B_{3}} \\
2A^{+} &   -(2A_{3}+K)&\frac{2A^{+}B^{-}}{1+B_{3}} \\
\frac{2(1+A_{3})B^{+}}{1+B_{3}}   & \frac{2A^{-}B^{+}}{1+B_{3}} & K-2
\end{array} \right),\label{2.2} 
\end{eqnarray}
where
\begin{eqnarray}
K= \frac{(1+A_{3})(1-B_{3})}{1+B_{3}}.\label{2.2} 
\end{eqnarray}
This is the direct M-transformation. This M-transformation allows us  to find solutions of the $\Gamma$-spin system (6.14) if we know the solutions of the coupled M-LIII equation (2.10)-(2.11). 
\subsection{Inverse M-transformation}
According to the inverse M-transformation, solutions of the coupled M-LIII equation can be expressed by the components of the $\Gamma$-spin system as
\begin{eqnarray}
A =\frac{1}{1-\Gamma_{33}}
\left ( \begin{array}{cc}
\Gamma_{11}-\Gamma_{22}   & 2\Gamma_{12} \\
2\Gamma_{21} &   \Gamma_{22}-\Gamma_{11}\end{array}\right),\label{2.2} 
\end{eqnarray}
\begin{eqnarray}
B =\frac{1}{1-\Gamma_{22}}
\left ( \begin{array}{cc}
\Gamma_{11}-\Gamma_{33}   & 2\Gamma_{13} \\
2\Gamma_{31} &   \Gamma_{33}-\Gamma_{11}\end{array}\right).\label{2.2} 
\end{eqnarray}
The transformations (7.3)-(7.4)  is called the inverse  M-transformation. Using the inverse  M-transformation,  we can   find solutions of  the coupled M-LIII equation (2.10)-(2.11),  if we know the solutions of the $\Gamma$-spin system (6.14).

 \section{Conclusions}
 In this paper, we have shown how the dynamics of two interacting and  moving curves in some space $R^{n}$ can be related to the dynamics of the coupled spin systems, namely, the coupled M-LIII equation.  Next, after some algebra we have proved that these two interacting and moving  curves are  related with the Manakov system. On the other hand, it is well-known that the Manakov system is equivalent to the $\Gamma$-spin system. Also,  we have presented the transformations which established the relation between solutions of the $\Gamma$-spin system and the coupled M-LIII equation. Our results can also be generalized to higher dimensional spaces (see e.g. refs. \cite{akbota1}-\cite{akbota8}). Work along these lines is in progress.


\begin{thebibliography}{99}

\bibitem{Gut} Gutshabash E.  Sh.  Zapiski nauchnyh
seminarov POMI,  {\bf 269}, 164-179  (2000)
\bibitem{Huang} 
Nian-Ning Huang, Bing Xu.  Commun. Theor. Phys.,   {\bf 12}, 121-126 1989).
\bibitem{R13} G.  Nugmanova, Z. Zhunussova, K. Yesmakhanova, G. Mamyrbekova, R. Myrzakulov.   International Journal of Mathematical, Computational, Statistical, Natural and Physical Engineering, {\bf 9}, N8, 328-331 (2015). 
\bibitem{He1} J.-S. He, Y. Cheng, Y.-S. Li. Commun. Theor. Phys.,   {\bf 38}, 493-496 (2002).
\bibitem{Saleem} U. Saleem, M. Hasan.  J. Phys. A: Math. Theor., {\bf  43}, 045204 (2010). 

\bibitem{royal}
M. Lakshmanan, Phil. Trans. R. Soc. A, {\bf 369} 1280-1300 (2011).


\bibitem{l77}
M. Lakshmanan, Phys. Lett. A, {\bf 64}, 53-54 (1977).

\bibitem{t77}
L.A. Takhtajan, Phys. Lett. A,  {\bf 64}, 235-238 (1977).

\bibitem{senthilkumar}
C. Senthilkumar, M. Lakshmanan, B. Grammaticos, A. Ramani, Phys. Lett. A {\bf 356} 339-345 (2006).

\bibitem{ishimori}
Y. Ishimori, Prog. Theor. Phys. {\bf 72} 33 (1984).

\bibitem{myrzakulov-391}	R. Myrzakulov, S. Vijayalakshmi, G. Nugmanova , M. Lakshmanan 		Physics Letters A, {\bf 233
}, 14-6,  391-396 (1997). 	

\bibitem{myrzakulov-2122}	R. Myrzakulov, S. Vijayalakshmi, R. Syzdykova, M. Lakshmanan, J. Math. Phys., {\bf 39},   2122-2139 (1998). 	

\bibitem{myrzakulov-3765}		R. Myrzakulov, M. Lakshmanan, S. Vijayalakshmi, A. Danlybaeva  , J. Math. Phys., {\bf 39},  3765-3771 (1998). 	



\bibitem{MK7}		Myrzakulov R, Danlybaeva A.K, Nugmanova G.N.  
		Theoretical and Mathematical Physics, 
V.118, 13, P. 441-451 (1999).


\bibitem{myrzakulov-9535}	Myrzakulov R., Nugmanova G., Syzdykova R.	 	Journal of Physics A: Mathematical \& Theoretical, V.31, 147, P.9535-9545 (1998). 	

\bibitem{myrzakulov-715}		Myrzakulov R., Daniel M., 
Amuda R.  	
	Physica A., V.234, 13-4, P.715-724 (1997). 	
\bibitem{myrzakulov-83}	Myrzakulov R., Makhankov V.G.,  Pashaev O.Ê.		Letters in Mathematical Physics, V.16, N1, P.83-92 (1989)	
\bibitem{myrzakulov-233}		Myrzakulov R., Makhankov V.G., Makhankov A.  		Physica Scripta, V.35, N3, P. 233-237 (1987)	
\bibitem{myrzakulov-378}		Myrzakulov R., Pashaev O.Ê., Kholmurodov Kh.		Physica Scripta, V.33, N4, P. 378-384 (1986) 
\bibitem{myrzakulov-1576}	Anco S.C.,
Myrzakulov R.		Journal of Geometry and Physics, v.60, 1576-1603 (2010)

\bibitem{myrzakulov-543} Myrzakulov R., 
Rahimov F.K., Myrzakul K., Serikbaev N.S.	\textit{On the geometry of stationary Heisenberg ferromagnets}. In: "Non-linear waves: Classical and Quantum Aspects", Kluwer Academic Publishers, Dordrecht, Netherlands, P. 543-549 (2004)
\bibitem{myrzakulov-535} Myrzakulov R., 
Serikbaev N.S., Myrzakul Kur., Rahimov F.K.	\textit{On continuous limits of some generalized compressible Heisenberg spin chains}. 	Journal of NATO Science Series II.  Mathematics, Physics and Chemistry, V 153, P. 535-542 (2004)
\bibitem{R14} R.Myrzakulov, G. K. Mamyrbekova, G. N. Nugmanova, M. Lakshmanan.  Symmetry, {\bf 7}(3), 1352-1375 (2015). [arXiv:1305.0098]
\bibitem{R15} R.Myrzakulov, G. K. Mamyrbekova, G. N. Nugmanova, K. Yesmakhanova,  M. Lakshmanan.  	Physics Letters A, {\bf 378}, N30-31, 2118-2123 (2014). [arXiv:1404.2088]
\bibitem{myrzakulov-248} Myrzakulov R., 
Martina L., Kozhamkulov T.A.,
Myrzakul Kur.	\textit{Integrable Heisenberg ferromagnets and soliton geometry of curves and surfaces}.	In book: "Nonlinear Physics: Theory and Experiment. II".  World Scientific, London, P. 248-253 (2003)
\bibitem{myrzakulov-314} Myrzakulov R.	\textit{Integrability of the Gauss-Codazzi-Mainardi equation in 2+1 dimensions}.	In  "Mathematical Problems of Nonlinear Dynamics", Proc. of the Int. Conf. "Progress in   Nonlinear sciences", Nizhny Novgorod, Russia, July 2-6, 2001, V.1, P.314-319 (2001)
\bibitem{Chen1} Chen Chi, Zhou Zi-Xiang. \textit{Darboux Tranformation and Exact Solutions of the Myrzakulov-I Equations}. Chin. Phys. Lett., {\bf 26}, N8, 080504 (2009)
\bibitem{Chen2} Chen Hai, Zhou Zi-Xiang. \textit{Darboux Transformation with a Double Spectral Parameter for the Myrzakulov-I Equation}. Chin. Phys. Lett., {\bf 31}, N12, 120504 (2014)
\bibitem{Zhao} Zhao-Wen  Yan, Min-Ru Chen, Ke Wu, Wei-Zhong Zhao. J. Phys. Soc. Jpn., {\bf 81}, 094006 (2012)
\bibitem{Yan} Yan Zhao-Wen, Chen Min-Ru, Wu Ke, Zhao Wei-Zhong. Commun. Theor. Phys., {\bf 58}, 463-468 (2012)
\bibitem{Es} K.R. Ysmakhanova, G.N. Nugmanova, Wei-Zhong Zhao, Ke Wu. \textit{Integrable inhomogeneous Lakshmanan-Myrzakulov equation}, [nlin/0604034]
\bibitem{Zhen-Huan} Zhen-Huan Zhang, Ming Deng, Wei-Zhong Zhao, Ke Wu. \textit{On the integrable inhomogeneous Myrzakulov-I equation}, [arXiv: nlin/0603069]
\bibitem{myrzakulov-1397} 
Martina L,  Myrzakul Kur.,  Myrzakulov R, Soliani G.	Journal of Mathematical  Physics, V.42, 13, P.1397-1417 (2001).
\bibitem{Wu} Xiao-Yu Wu,  Bo Tian, Hui-Ling Zhen, Wen-Rong Sun and Ya Sun. Journal of Modern Optics, 2015. 
\bibitem{M35} Z.S. Yersultanova, M. Zhassybayeva, K. Yesmakhanova,  G. Nugmanova,   R. Myrzakulov.   International Journal of Geometric Methods in Modern Physics,  {\bf 13}, N1, 1550134 (2016).. [arXiv:1404.2270]

\bibitem{Bordag1} Bordag L.A., Yanovski A.B. J. Phys. A: Math. Gen., {\bf 28}, 4007-4013 (1995)
\bibitem{Bordag2}Bordag L.A., Yanovski A.B. J. Phys. A: Math. Gen., {\bf 29}, 5575-5590 (1996)
\bibitem{Kostov} N. A. Kostov,  R. Dandoloff,  V. S. Gerdjikov,  G. G. Grahovski. \textit{The Manakov system as two moving interacting curves},  In the Proceedings of the International Workshop "Complex structures and vector fields", August 21--26, 2006, Sofia, Bulgaria. Eds.: K. Sekigawa, S. Dimiev. World Scientific (2007)
\bibitem{akbota1} Myrzakul  Akbota. \textit{Integrability of two coupled curves and geometrically equivalent spin analogue of the Manakov equation}. Vestnik ENU, N2, 95-99 (2016)  
\bibitem{akbota2} Myrzakul  Akbota. \textit{Equivalence between the coupled M-LIII equation and the $\Gamma$-spin system}. Vestnik ENU,  N3, 89-94  (2016) 
\bibitem{akbota5} Myrzakul  Akbota. \textit{Gauge equivalence between the coupled M-LIII equation and the Manakov  system}. Vestnik ENU,  N1, 54-60  (2016)
\bibitem{akbota3} Myrzakul  Akbota and Myrzakulov Ratbay. \textit{Motion of two interacting curves and surfaces: integrable reductions and soliton equations}. Vestnik ENU,  N4, 45-57  (2014)
\bibitem{akbota4} Myrzakul  Akbota and Myrzakulov Ratbay. \textit{Integrable Motion of Two Interacting Curves and Heisenberg Ferromagnetic Equations}, Abstracts of XVIII-th Intern. Conference "Geometry, Integrability and Quantization", June 3-8, 2016, Bulgaria.
\bibitem{akbota6} Myrzakul  Akbota and Myrzakulov Ratbay. \textit{Integrable motion of two interacting curves, spin systems and the Manakov system}, [arXiv:1606.06598]
\bibitem{akbota7} Myrzakul  Akbota and Myrzakulov Ratbay. \textit{Darboux transformations and exact soliton solutions of integrable coupled spin systems related with the Manakov system}, [arXiv:1607.08151]
\bibitem{akbota8} Myrzakul  Akbota and Myrzakulov Ratbay. \textit{Integrable geometric flows of interacting curves/surfaces, multilayer spin systems and the vector nonlinear Schrodinger equation}, [arXiv:1608.08553]\end{thebibliography}
 \end{document}